# Light Polarization Sensitive Transistor Action in the van der Waals ferroelectric 4H-$SnS_2$

*Abhishek Bajgain, Rabindra Basnet, Subhashree Chatterjee, Alexander Samokhvalov, and Ramesh C. Budhani*[*]

Abhishek Bajgain, Rabindra Basnet, Subhashree Chatterjee, Alexander Samokhvalov, Ramesh C. Budhani
Department of Physics
Morgan State University
Baltimore, MD, 21251, USA
E-mail: ramesh.budhani@morgan.edu

**Funding:** This work at Morgan State University has been funded by the Department of Defense through grant # W911NF2120213.

**Van der Waals (vdW) ferroelectric semiconductors provide a unique platform for exploring the interplay between spontaneous polarization, electronic transport, ion migration, and photocarrier generation at the nanoscale. Here, we establish a room-temperature ferroelectric state in $SnS_2$ and its sensitivity to structural polytype by directly contrasting the centrosymmetric 2H-phase with the polar 4H-phase. Raman spectroscopy distinguishes**

**the 2H and 4H polymorphs of $SnS_2$ through their characteristic phonon fingerprints. Piezoresponse force microscopy confirms room-temperature ferroelectricity in the 4H phase, while the 2H phase exhibits no ferroelectric response. The built-in polarization in a three-terminal transistor device of the 4H-$SnS_2$ is modulated significantly by electrostatic gating and on exposure to linear and circularly polarized light. This device reveals polarization-controlled output characteristics with distinct gate-voltage induced hysteretic response and thermally activated carrier transport, confirming *p*-type semiconducting behavior strongly coupled to ferroelectric polarization. The photoresponse likewise exhibits polarization-assisted carrier separation, sublinear power-law scaling, a nonmonotonic temperature response correlated with the characteristic Raman modes, ferroelectric hysteresis, and a pronounced dependence on the circular and linear polarization states of the incident laser beam. These results establish 4H-$SnS_2$ as a promising material system for polarization-driven electronic and optoelectronic technologies, including nonvolatile memory and photoferroelectric functionalities.**

## 1. Introduction

The interplay of multiple order parameters in solids can give rise to exotic photo-electronic phenomena by intertwining distinct symmetry-breaking mechanisms and collective degrees of freedom. In layered ferroelectric semiconductors, the intimate coupling between ferroelectric polarization, electronic band structure, and crystal symmetry provides a unique pathway to actively control electronic states using the internal switchable electric field[1,2]. This enables prominent electronic and optoelectronic responses, including built-in polarization-dependent carrier transport[3], nonlinear optical effects[4] which are sensitive to the polarization of the light field[5,6] and electrically tunable photoresponse[7]. Moreover, the reduced dimensionality arising from van der Waals (vdW)-type stacking in these materials amplifies the coupling between polarization and electronic states by weakening interlayer screening and enhancing structural flexibility. As a result, layered ferroelectrics exhibit thickness-dependent electronic transitions[8] and tunable ferroelectric order down to the two-dimensional (2D) limit[9,10]. Further, their 2D layered structure allows the unconventional "sliding ferroelectricity", where a spontaneous polarization emerges from in-plane relative displacements between otherwise centrosymmetric monolayers[11]. The resulting inversion-symmetry breaking can also enable light polarization-dependent photo responses, including circular-polarization-sensitive photocurrent[12,13]. These characteristics establish layered ferroelectric semiconductors as appealing platforms for electrically reconfigurable electronic and optoelectronic functionalities.

There has been a surge in understanding ferroelectricity in vdW materials, as evident from the recent works on $Cu(In,Cr)P_2S_6$[9,14,15], $In_2Se_3$[16], $Sn_2P_2(S,Se)_6$[17,18], $SnP_2S_6$[19], $Bi_2O_2Se$[20,21], $MoTe_2$[22], etc., and the Moiŕe structures thereof[23,24]. These systems also provide versatile platforms for engineering ferroelectricity via external perturbations such as strain, pressure, and interlayer coupling[25,26]. The net polarization in these materials generates an internal electrostatic field that modulates carrier density and band bending, thereby also influencing the photoconductivity by affecting photocarrier separation, recombination dynamics, and photoresponsivity[15]. While such effects are well established in conventional 3D ferroelectrics[27,28], the role of nanoscale ferroelectric order in 2D materials, particularly under external electrical or optical stimuli, is an emergent field of research that warrants detailed investigation.

In this context, ferroelectric and photonic control of charge transport represents a key research direction for advancing the mechanism of light-assisted photocarrier generation in vdW materials and its translation into potential opto-electronic devices. Unlike most well-known 2D ferroelectrics[9,14,17–21], $SnS_2$ is a simple two-element indirect band gap semiconductor from the bulk to the monolayer limit[29]. This system provides a versatile platform for nanoscale devices, including field-effect transistors [29] and neuromorphic circuits[30]. Nevertheless, ferroelectricity has not been conclusively established in pristine $SnS_2$. This is likely due to the coexistence of multiple structural polytypes with distinct stacking symmetries and associated electronic properties[31]. In particular, a recent theoretical study predicted spontaneous out-of-plane ferroelectric polarization in the hexagonal 4H-$SnS_2$ ($P6_3mc$)[31], whereas the commonly observed 2H polymorph (trigonal $P\bar{3}m1$) possesses a centrosymmetric nonpolar crystal structure and therefore lacks a net ferroelectric dipole moment[29,32,33]. Only recently, some indications of signatures of out-of-plane ferroelectricity were reported in a few-layer $SnS_2$ nanobelt[34]. However, the role of crystal symmetry in stabilizing ferroelectricity remains to be fully clarified in $SnS_2$. The observed ferroelectric polarization in this work[34] was attributed to intercalation-induced stress/strain and the resulting microstructural modifications, including vacancy formation, intralayer expansion, and interlayer sliding, rather than to an intrinsic crystal-symmetry-driven origin.

Here, we systematically compare the 2H and 4H polymorphs of $SnS_2$ single crystals to uncover the role of crystal symmetry in stabilizing room-temperature ferroelectricity. We demonstrate that the polar 4H-phase hosts ferroelectric order at ambient conditions, and its spontaneous polarization strongly modulates the electrical transport and light-responsive transistor characteristics. The crystal polytype of $SnS_2$ is identified by Raman spectroscopy, where distinct Raman modes clearly distinguish the 2H and 4H phases. This work demonstrates that the 4H phase of $SnS_2$ hosts room-temperature ferroelectricity, whereas the 2H polymorph remains non-ferroelectric. The ferroelectric ordering in 4H-$SnS_2$ strongly modulates the electrical transport and photocarrier dynamics, displaying gate-tunable hysteresis, laser power, and polarization state (linear and circular) dependent photoconductivity, and a distinct temperature dependence of the transistor characteristics. In particular, the correlated evolution of Raman modes, ferroelectric hysteresis, and photocurrent reveals strong coupling between lattice symmetry, polarization, and charge dynamics. Clearly, this discovery of ferroelectricity in 4H-$SnS_2$ establishes it as a promising platform for polarization-controlled electronics and optoelectronics, including nonvolatile memory. Further, unlike many other complex 2D ferroelectrics, $SnS_2$ is a simple non-reactive binary compound offering ease of device fabrication.

## 2. Results and Discussion

### 2.1. Ferroelectric phase Identification in 2H and 4H $SnS_2$ Polymorphs

The CVT growth of $SnS_2$ yielded two distinct sets of crystals. Notably, as seen in **Figure** S1, both G and R crystals were simultaneously obtained within the same growth ampoule near the high-temperature (source) and low-temperature (sink) zones, respectively. The preferential growth of these crystals in distinct thermal regions highlights the strong temperature dependence of phase formation in $SnS_2$. This behavior suggests that subtle variations in local thermodynamic growth conditions play a decisive role in selecting and stabilizing the competing 2H and 4H polymorphs. Such pronounced sensitivity to growth conditions complicates the synthesis, characterization, and distinction of individual polymorphs of $SnS_2$. Their distinct colors further suggest differences in the optical gap. However, the elemental analyses by EDS (**Figures** 1a and S2a) reveal a nearly identical Sn:S $\approx$ 1:2 stoichiometry for both sets of crystals. Compositional homogeneity in these crystals was verified by EDS measurements at multiple points together with large-area elemental mapping, as shown in **Figures** 1a and S2a. The XRD patterns of both crystals exhibit similar (00*L*) reflections with only a slight shift in peak position of $\Delta(2\theta) \approx 0.4°$ (**Figures** 1b and S2b), indicating closely related layered crystal structures.

Raman spectroscopy was used to identify the structure polytype of these $SnS_2$ crystals. As shown in **Figure** 1(c), the room-temperature Raman measurement of the wine-red crystal reveals a prominent mode near 313.4 $cm^{-1}$ and a doublet peak at ~202.4 and ~212.8 $cm^{-1}$ (inset). In contrast, the light-green crystal exhibits a similarly sharp mode near 313.9 $cm^{-1}$ but lacks peak splitting in the low-energy mode at ~204.6 $cm^{-1}$ (**Figure** S2c). These spectra have been analyzed in the framework of reported Raman results of $SnS_2$ polytypes[29,32]. Here, the dominant high-energy peak is attributed to a mixed $A_{1g}$ and $E_g$ modes, which closely matches the $A_{1g}$ mode of 2H-$SnS_2$ and is therefore not suitable for polytype distinction. The low-energy $E_g$ mode (~207 $cm^{-1}$), on the other hand, enables a clear phase differentiation because in 2H-$SnS_2$ it is a single intense line, whereas for 4H-$SnS_2$ it has a characteristic double structure[29,32], as seen in the inset of **Figure** 1(c). This $E_g$ mode corresponds predominantly to in-plane vibrations of the S atoms within the basal plane[32]. The observed doublet structure (peaks A and B) likely arises from symmetry lowering and stacking-induced splitting in the 4H phase[32]. On the other hand, the higher-frequency mode near ~313 $cm^{-1}$, labeled as $A_{1g} + E_g$, is primarily associated with out-of-plane vibrations of the S atoms along the *c*-axis, with additional E-type admixture due to reduced symmetry and interlayer coupling[32].

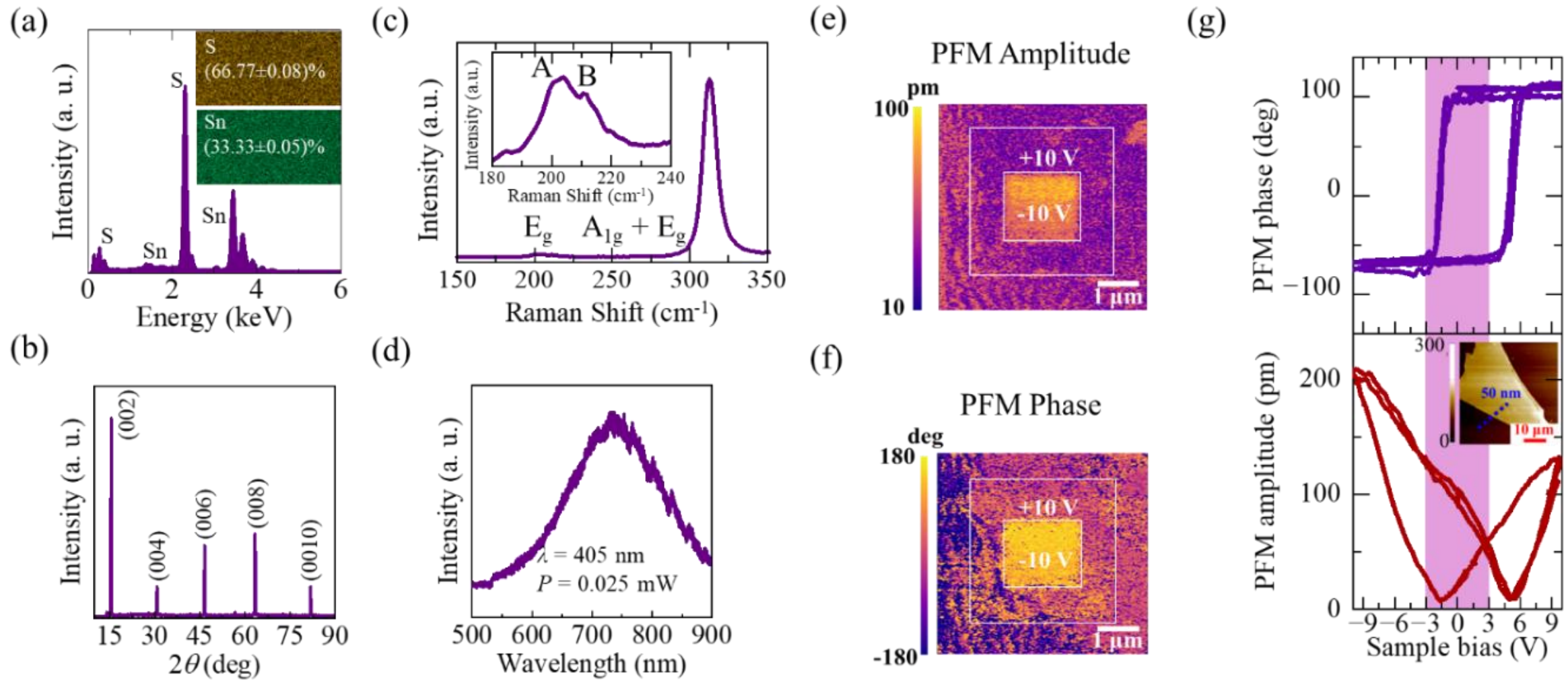

***Figure 1. Structural, optical, and ferroelectric characterization of 4H-SnS₂.*** *(a) EDS spectrum of a CVT-grown 4H-SnS₂ single crystal showing Sn and S peaks. The inset displays EDS elemental mapping of Sn and S over a larger area of the crystal. (b) X-ray diffraction (XRD) pattern of the same single crystal showing strong (00L) reflections, confirming preferential orientation along the c-axis. (c) Raman spectrum of 4H-SnS₂, highlighting two characteristic* $E_g$ *and* $A_{1g} + E_g$ *phonon modes, which corresponds to the in-plane and out-of-plane S vibrations, respectively. The inset highlights the doublet structure (peaks A and B) of the low-frequency* $E_g$ *mode. (d) Photoluminescence (PL) spectrum measured at room temperature under excitation with a 405 nm laser at a power of 0.025 mW, illustrating the intrinsic semiconducting behavior of 4H-SnS₂. (e) PFM amplitude and (f) phase images acquired after writing a box-in-box pattern using a ±10 V DC bias. (g) Off–field PFM phase hysteresis loop (upper panel) and corresponding amplitude loop (lower panel) recorded during the DC bias switching process for 50 nm thick flake shown in the inset.*

The 2H polytype of $SnS_2$ consists of two sulfur layers per unit cell, forming a centrosymmetric stacking. In contrast, the 4H hexagonal polytype contains four sulfur layers per unit cell with a more complex stacking sequence involving alternating Sn and S layer arrangements, leading to inversion symmetry breaking[31,33]. Consistent with this, the XRD patterns yield a *c*-axis lattice parameter of ~5.86 Å and ~11.42 Å for 2H- and 4H-$SnS_2$ crystals, respectively. An earlier theoretical study predicted a net out-of-plane ferroelectric polarization in hexagonal 4H-$SnS_2$[31], analogous to 2D ferroelectric material $CuInP_2S_6$[14]. To confirm room-temperature ferroelectricity in $SnS_2$, piezoresponse force microscopy (PFM) measurements were performed on a thin (~50 nm) exfoliated flake (**Figures** 1e-g). A box-in-box pattern written with ±10 V DC bias shows opposite phase contrasts in the inner and outer regions (**Figures** 1e and f), indicating reversible domain switching. Off-field PFM phase hysteresis exhibits a 180° phase reversal, while the corresponding butterfly-shaped amplitude loop shows minima near ~3 V (**Figure** 1g), characteristic of polarization switching. The finite offset of the hysteresis loop from zero bias likely originates from an internal built-in electric field due to asymmetric Schottky barriers at the top and bottom interfaces of the sample.

To establish the intrinsic ferroelectric nature of mechanically exfoliated 4H-$SnS_2$, we systematically investigated its domain switching behavior, polarization retention, switching reproducibility, and thickness-dependent response using PFM and Kelvin probe force microscopy (KPFM). As shown in **Figures** 2(a and b), local ferroelectric domains were written by applying ±9 V DC bias through a conductive AFM tip. The KPFM surface potential maps taken immediately after poling exhibit a pronounced contrast between oppositely polarized regions, originating from polarization-induced bound charges that modify the local surface potential. The corresponding surface topography (**Figure** S3, Supporting Information) exhibits no discernible morphological variations after domain writing, confirming that the observed contrast originates from polarization reversal rather than topographic artifacts[35]. Importantly, the written potential contrast remains clearly distinguishable after 48 h with only a slight reduction in magnitude, demonstrating long-lived polarization retention. The temporal stability of the electrostatic contrast indicates that the written states are governed by stable remanent polarization rather than rapidly relaxing surface charges or transient charge injection during the domain writing process. **Figures** 2(c and d) present the corresponding PFM phase and amplitude maps acquired from the same region examined by KPFM, revealing a clear 180° phase contrast and reversible polarization switching, characteristic of ferroelectric domain reversal. The stability of the switched ferroelectric domains was further visualized through time-dependent PFM imaging on a ~86 nm 4H-$SnS_2$ flake. A box-in-box domain structure written using ±9 V reversible DC biases remain clearly preserved after 15 h and even after 60 h (**Figures** 2e-g), with minimal degradation of the phase contrast or domain morphology. The absence of noticeable domain back-switching over extended periods demonstrates excellent retention of the remanent polarization and indicates that the switched domains are energetically stable at room temperature. The corresponding PFM amplitude images are shown in Supporting Information (**Figure** S4). A similar retention behavior is observed in a ~41 nm thick 4H-$SnS_2$ flake, where the PFM phase contrast remains essentially unchanged after 12 h, further demonstrating the stability of the remanent polarization (**Figure** S5, Supporting Information). Off-field switching spectroscopy was performed on flakes with thicknesses of approximately 38, 47, 86, and 110 nm (**Figures** 2h–k). All samples consistently exhibit robust phase hysteresis together with well-defined butterfly amplitude loops, confirming that reversible polarization switching is preserved across a broad thickness range. The repeatability of polarization switching was evaluated through repeated domain writing and erasing experiments. As shown in **Figure** 2l, square domains were alternately written using ±5 V biases for 20 consecutive cycles while maintaining reproducible phase contrast throughout the experiment. The absence of observable degradation after repeated switching demonstrates excellent endurance and reversible polarization switching in the exfoliated flakes. The corresponding PFM amplitude images after $1^{st}$, $10^{th}$, and $20^{th}$ writing-erasing cycles are shown in Supporting Information (**Figure** S6). To further confirm that the observed switching originates from intrinsic ferroelectricity rather than electrostatic artifacts, control experiments including frequency-dependent PFM loop measurements, comparison of on-field and off-field switching spectroscopy, and repeated switching spectroscopy over ten consecutive cycles were performed (Supporting Information: **Figures** S7-9). All these measurements exhibited nearly identical phase and amplitude hysteresis loops with negligible variation in coercive voltage, confirming stable, reproducible polarization switching without significant charge accumulation or fatigue. These observations are characteristic of a remanent electromechanical response arising from ferroelectric polarization and are inconsistent with transient electrostatic or electrochemical effects, which typically relax with time[14,21]. In the present study, the observed ferroelectric switching is primarily associated with the out-of-plane polarization component of the 4H-$SnS_2$ crystal, which was directly probed by vertical PFM. To further determine the polarization orientation, in-plane PFM measurements were performed on a 110 nm thick flake (**Figure** S10). Compared with the pronounced out-of-plane response (Figure 2k), the in-plane piezoresponse is considerably weaker, indicating that the spontaneous polarization is predominantly oriented along the out-of-plane direction. Although these PFM measurements provide compelling evidence for local ferroelectric switching, future macroscopic P-E and PUND measurements will be valuable for quantitatively determining the intrinsic polarization and coercive field. Nevertheless, these results collectively provide the first direct experimental evidence for theoretically predicted[31] ferroelectricity in the 4H-$SnS_2$ crystal. In contrast, similar PFM measurements on 2H-$SnS_2$ reveal no detectable ferroelectric response (**Figure** S2e).

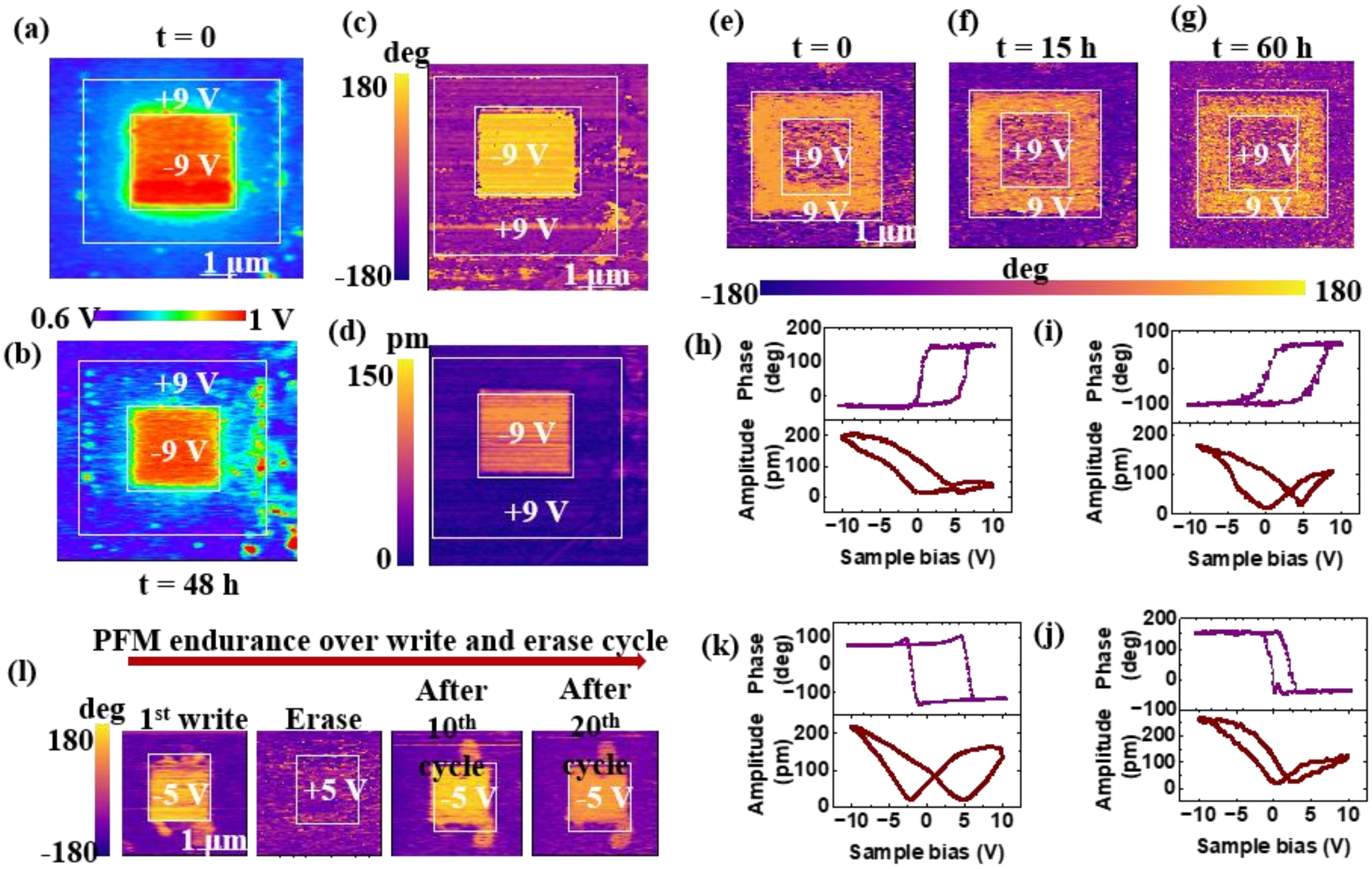

***Figure 2.*** *Out-of-plane ferroelectric characterization of 4H-$SnS_2$ nanoflakes. (a,b) Time-dependent evolution of the surface potential distribution measured by non-contact KPFM after local ferroelectric domain writing on a ~100 nm thick 4H-$SnS_2$ flake. Opposite polarization states were locally generated by applying ±9 V DC biases through the conductive AFM tip, and the corresponding surface potential contrast was recorded immediately after poling (t = 0) and after 48 h (t = 48 h). (c,d) Corresponding PFM phase and amplitude hysteresis loops obtained from the same region, showing characteristic polarization reversal and piezoresponse evolution under applied electric fields. (e–g) Time-dependent PFM phase imaging of a box-in-box domain pattern written on an ~86 nm thick 4H-$SnS_2$ flake using ±9 V DC bias. The written domains were monitored*

*immediately after writing (t = 0), after 15 h (t = 15 h), and after 60 h (t = 60 h). (h–k) Off-field switching spectroscopy PFM measurements showing phase (upper panels) and amplitude (lower panels) hysteresis loops acquired from 4H-$SnS_2$ flakes with different thicknesses of ~38, 47, 86, and 110 nm, respectively. (l) Ferroelectric endurance measurement performed on a ~110 nm thick 4H-$SnS_2$ flake by repeated domain writing and erasing cycles. Square domains were sequentially written and erased using ±5 V DC biases, and the reproducible phase contrast over 20 consecutive switching cycles demonstrates the reliable and repeatable polarization reversal capability of 4H-$SnS_2$ nanoflakes.*

### 2.2. Ferroelectric control of charge transport in 4H-$SnS_2$

We focus on the 4H-$SnS_2$ sample to further investigate the influence of built-in ferroelectric polarization on its electrical transport and photoresponse. The 4H phase exhibits a broad PL emission spanning ~560-880 nm (2.21 – 1.41 eV) (**Figure** 1d). In contrast, 2H-$SnS_2$ shows a comparatively higher-energy PL response extending over ~ 480-700 nm (2.58 – 1.77 eV) (**Figure** S2d), indicating its larger optical band gap compared to the 4H phase. Furthermore, the PL spectrum of 2H-$SnS_2$ also features oscillatory behavior, which may originate from excitonic transitions and/or optical interference effects associated with multiple internal reflections of the emitted PL within the layered flakes.

In ferroelectric semiconductors, charge transport arises from the combined contributions of mobile carriers and dynamic polarization[21,36]. The ferroelectric effect in 4H-$SnS_2$ crystal is demonstrated by the room-temperature output characteristics of the Field-Effect Transistor (FET). The schematic of the experimental setup and $SnS_2$ FET, showing the source, drain, and gate contacts, are presented in **Figures** 3(a) and 3(b), respectively. Unlike previous device architectures employing a separate vdW dielectric layer, our all-$SnS_2$ transistor integrates the semiconducting channel and ferroelectric gate dielectric within the same single crystal. The use of van der Waals (vdW) dielectrics as gate materials has recently emerged as an effective strategy for two-dimensional electronics because they provide efficient electrostatic control while preserving the intrinsic properties of layered materials[37,38]. Recent studies have demonstrated that layered dielectrics with moderate-to-high dielectric constants can effectively work as gate dielectric in two-dimensional field-effect transistors. The out-of-plane dielectric constant of 4H-$SnS_2$ is $\varepsilon_r$ = 6.98[39], which is comparable to or larger than that of several reported vdW dielectric materials such as $CaI_2$ ($\varepsilon_r$ = 3.5), $MgCl_2$ ($\varepsilon_r$ = 2.7), etc[37,38]. The layered crystal structure of wide-bandgap vdW semiconductors like 4H-$SnS_2$ [$E_g \approx 2.31$ eV; details in the Supplementary Note 4 and **Figure** S11] gives rise to pronounced transport anisotropy due to the weak van der Waals coupling between adjacent layers, resulting in an out-of-plane resistivity a few orders of magnitude larger than the in-plane resistivity[40]. Consequently, a significant vertical potential drop develops across the vdW layers, suppressing carrier injection toward the gate and confining charge transport predominantly within the upper layers of the channel, thereby minimizing gate leakage and enabling efficient electrostatic gate control. This behavior is confirmed by our in-plane and out-of-plane transport measurements, which show that the out-of-plane resistivity is approximately 142-250 times larger than the in-plane resistivity over the measured temperature range, yielding a resistivity ratio of $\rho_{in}/\rho_{out} \approx$ 0.4-0.7% (**Figure** S12). Based on this anisotropy, we further analyzed the current distribution using a resistor-network model (details in the Supplementary Note S6), where the intra-layer and inter-layer transport are represented by resistances Rp and Rs, respectively (**Figure** S14). The analysis shows that for a transport anisotropy ratio of ~142 at room temperature, more than 99.476% of the injected source-drain current remains confined to the directly contacted in-plane channel, whereas less than 0.6% is redistributed vertically through the van der Waals gaps (**Table** S1 and **Figure** S15). Consequently, the vertical gate leakage is expected to remain below approximately 1% of the lateral source-drain current, while efficient in-plane conduction is maintained between the source and drain electrodes. Our resistor-network model is qualitatively consistent with an earlier numerical simulation[40], which demonstrated an approximately 50% reduction in drain current as the channel thickness increased from 4 to 16 nm for a comparable transport-anisotropy factor. In both models, the reduction arises because the large cross-plane resistance suppresses carrier injection from the top contacts into deeper conducting layers near the gate. These results demonstrate that strong transport anisotropy effectively suppresses cross-plane leakage while preserving electrostatic field penetration, enabling the bulk 4H-$SnS_2$ crystal to simultaneously function as both the semiconducting channel and the gate dielectric in our device architecture.

The $I_d$-$V_d$ characteristics (-5 to +5 V) measured under different programming voltages ($V_p$) ranging from −10 V to +10 V in steps of 2V show nearly linear and symmetric behavior as the $V_p$ changes from +10 V to −4 V, indicating an ohmic contact. When $V_p$ is further changed from −6 V to −10 V, the $I_d$-$V_d$ characteristics exhibit ohmic conduction only in the low-$V_d$ regime between −2

V and +2 V (blue rectangle in **Figure 3c**). Beyond this range, pronounced nonlinearity and asymmetry emerge, with significantly higher $I_d$ under negative $V_d$ compared to the positive $V_d$. The inset of **Figure 3c** shows the extracted $I_d$ as a function of $V_p$ at fixed $V_d$ = +5V (red curve) and $V_d$ = −5V (black curve). The drain current exhibits weak dependence on $V_p$ at $V_d$ = +5V. On the other hand, for −5V $V_d$ it remains nearly constant at $I_d \approx -15$ nA from $V_p$ = +10V to −4V, followed by a sharp increase to $I_d \approx -30$ nA for $V_p$ between −6 V and −10 V. This behavior indicates that the 4H-$SnS_2$ FET is *p*-type in nature, with negative polarization promoting hole accumulation in the channel.

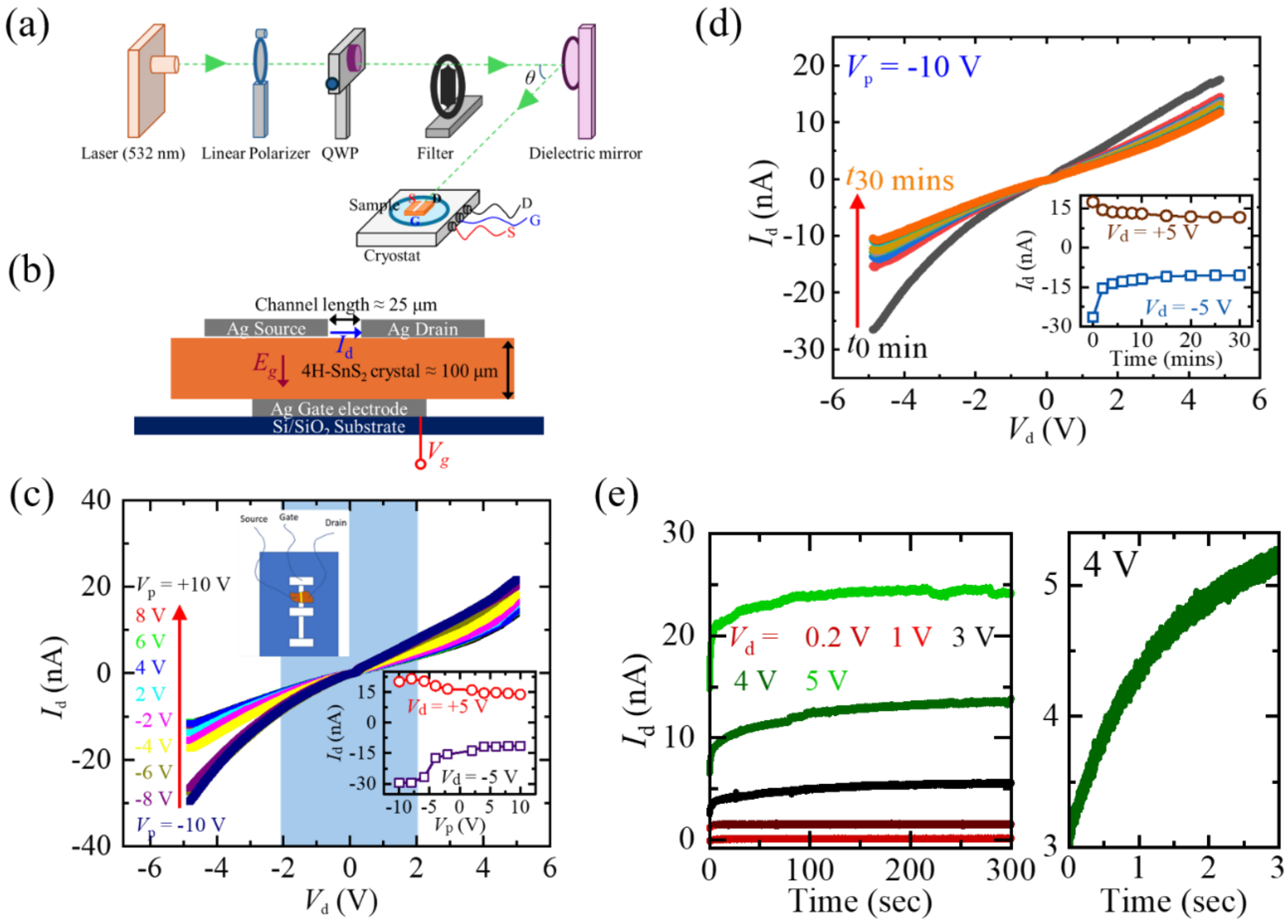

***Figure 3. Ferroelectric field-effect transistor characteristics of 4H-SnS₂.*** *(a) Schematic illustration of the electrical and photoresponse measurement setup. All measurements were performed under high-vacuum conditions inside a variable-temperature cryostat. (b) Cross-sectional schematic of the all-4H-$SnS_2$ three-terminal phototransistor. (c) Output characteristics ($I_d$-$V_d$) measured at room temperature under different programming voltages $V_p$ ranging from +10 V to −10 V at a step of 2V. The inset shows a schematic of the 4H-$SnS_2$ ferroelectric-FET device geometry with source, drain, and gate electrodes. The shaded region highlights the low-$V_d$ regime*

*where the response remains nearly linear. The extracted $I_d$ as a function of $V_p$ at fixed $V_d$ = +5V and −5 V is shown in the inset, revealing pronounced asymmetry consistent with p-type transport and ferroelectric polarization effects. (d) Time evolution of the $I_d$-$V_d$ characteristics measured at a fixed programming voltage of $V_p$ = -10 V over a duration of 20 mins, demonstrating the rapid decay of polarization-induced current enhancement due to carrier screening. The inset shows the corresponding time-dependent drain current at $V_d$ = ±5 V, highlighting the transient nature of polarization-assisted transport in 4H-$SnS_2$. (e) Left panel: The time-dependent drain current ($I_d$-t) measurements under fixed drain voltage ($V_d$ = 0.2 to 5 V). Right panel. The $I_d$-t plot for $V_d$ = 4 V.*

The pronounced asymmetry in the $I_d$-$V_d$ curves might be attributed to bound polarization charges[15,41]. It creates an internal field that favors hole flow under negative $V_d$ while partially opposing it under positive $V_d$, implying drain current modulation by the ferroelectric state of the channel. Notably, the onset of these $I_d$-$V_d$ characteristics at programming voltages more negative than −4 V is consistent with the threshold for ferroelectric polarization switching of ~3 V obtained from the PFM measurements (**Figure** 1g). To further confirm the role of polarization in 4H-$SnS_2$, $I_d$-$V_d$ measurements were performed at a constant programming voltage of −10 V and monitored over 30 minutes time (**Figure** 3d). The results show that the polarization-induced enhancement of $I_d$ to ~30 nA at $V_d$ = −5V persists only briefly (≈2-3 minutes) before gradually diminishing, whereas $I_d$ at $V_d$ = +5V remains nearly unchanged over the entire measurement duration (inset, **Figure** 3d). Consequently, after approximately 2-3 minutes, the FET output gradually evolves toward nearly linear $I_d$-$V_d$ characteristics over the entire drain-voltage range. This evolution occurs on a timescale several orders of magnitude longer than the intrinsic ferroelectric switching time, which typically ranges from sub-microseconds to milliseconds[42,43]. Therefore, the slow evolution implies the involvement of secondary relaxation processes, including carrier redistribution, polarization screening, charge trapping/detrapping, and possible Joule-heating effects[44]. To further distinguish these contributions, we performed time-dependent drain-current ($I_d$-$t$) measurements under fixed drain voltages $V_d$ = 0.2 to 5 V (left panel, **Figure** 3e). At low bias $V_d$ = 0.2-2 V, $I_d$ remains nearly constant throughout the measurement, indicating ohmic transport in the absence of polarization switching. In contrast, for $V_d$ ≥ 3 V, corresponding closely to the threshold for ferroelectric polarization switching inferred from the PFM measurements, the drain current exhibits an abrupt increase immediately after bias application, followed by a gradual relaxation to a steady-state value within a few seconds (right panel, **Figure** 3e). The coincidence between the onset of the current enhancement and the PFM polarization reversal strongly suggests that the initial current increase is triggered by polarization-induced modulation of the channel carrier density. The threshold electric field for the onset of ferroelectric switching is ≈ 1.2 kV/cm, which is much lower than that of $CuInP_2S_6$ (~25-30 kV/cm )[45], implying relatively easier polarization switching in 4H-$SnS_2$. The subsequent second-scale relaxation is likely described by carrier screening and charge trapping/detrapping, which progressively compensate for the polarization-bound charges and reduce the effective internal electric field. Importantly, the current does not decay back to its initial value but instead reaches a stable, bias-dependent nonzero plateau that persists throughout the measurement. The magnitude of this steady-state current increases systematically with drain bias (left panel; **Figure** 3e), indicating that only the transient component of the polarization-enhanced current is suppressed by carrier screening, while a substantial remanent current persists. The retention of this stable current demonstrates that the remanent ferroelectric polarization continues to modulate the channel conductivity even after partial screening of the internal electric field. This behavior is reminiscent of the ferroelectric transport mechanism in 2D ferroelectric transistors, where large electric fields give rise to a stable, nonvolatile conducting state with long memory retention[46,47]. This conclusion is strongly supported by the time-dependent KPFM and PFM measurements depicted in **Figure** 2.

Owing to the semiconducting nature of $SnS_2$, the channel hosts mobile carriers depending on the positions of the band edges relative to the Fermi level ($E_F$). Such carriers progressively screen the ferroelectric polarization, giving rise to a temporal decay of the polarization-induced effects–a well-known phenomenon in ferroelectric semiconductors[41,48–50]. In this process, free carriers redistribute to compensate polarization-induced bound charges, thereby reducing the effective internal electric field over time. The time-dependent relaxation measurements were conducted by applying the 100 ms programming pulse ($V_p$ = −10 −5, +5, and +10 V). The details of the measurements are discussed in the "Experimental" section. Following removal of the programming pulse, the drain current was monitored at a constant drain bias of $V_d$ = 3 V (**Figure** S16). Four distinct current states are obtained according to the programming voltage and remain well separated throughout the 300 s measurement. A slight initial relaxation is observed during the first few seconds, followed by a nearly constant steady-state current. The higher current after negative programming pulses is consistent with polarization-induced hole accumulation in the *p*- type 4H-$SnS_2$ channel, whereas positive programming pulses produce a lower-conductance state. The persistence of these programmed current levels demonstrates stable retention of the polarization-modulated channel conductance. The small initial relaxation is attributed to partial screening of the polarization-bound charges by mobile carriers, while the long-lived current plateau reflects the remanent ferroelectric polarization. These findings confirm that charge transport in $SnS_2$ is governed by a synergistic interplay between ferroelectric polarization and semiconductor carriers.

The mobile carriers create a non-uniform electric field that governs polarization switching. This coupling is also reflected in the gate-dependent transfer characteristics, where gate voltage ($V_g$) modulates the effective field in the channel. Using the parallel-plate approximation, the gate capacitance per unit area is given by, $\frac{C_g}{A} = \frac{\varepsilon_0 \varepsilon_r}{t}$. Here, $\varepsilon_r$ = 6.98 and thickness of the crystal ($t$) = 100 μm yields $\frac{C_g}{A}$ = 0.0618 nF cm$^{-2}$. Further, the effective gate field is estimated as $E_g = \frac{V_g}{t}$ , where $V_g$ is the applied gate voltage. For $V_g$ = 15 V and $t$ = 100 μm, the effective gate field is $E_g$ = 1.5 kVcm$^{-1}$. This field is comparable to the experimentally observed coercive field (≈ 1.2 kVcm$^{-1}$) for ferroelectric switching in 4H-$SnS_2$, as discussed above, indicating that the applied gate bias is sufficient to influence the ferroelectric polarization and channel conductivity[46], thereby enabling effective gate coupling. **Figure 4**(a) presents the $I_d$ as a function of $V_g$ measured from $T$ = 80 to 300 K at a fixed $V_d$ = +3 V. At room temperature (300 K), the $I_d$ is significantly higher under negative $V_g$ while it is strongly suppressed at positive $V_g$. The pronounced asymmetry in the transfer curves indicates that charge transport is dominated by holes, in agreement with the behavior observed in the programming voltage-dependent output characteristics presented in **Figure 3**. This effect may originate from gate-induced band bending, where negative $V_g$ drives the valence band above the $E_F$[19]. **Figure** S13 compares the gate-voltage dependence of the source-drain and source-gate current measured for the 4H-$SnS_2$ FeFET device. These measurements demonstrate that source-gate current remains nearly three orders of magnitude smaller than source-drain current, indicating negligible gate leakage, which is consistent with our mathematical model discussed in the Supplementary Note 6. The thermal response under electrostatic gating is elucidated by the evolution of $I_d$ as a function of temperature at fixed gate voltages from +15 V to −15 V (**Figure** 4b). At temperatures 80 to 220 K, $I_d$ exhibits weak dependence on both temperature and gate voltage. Above $T \geq 240$ K, $I_d$ remains nearly constant for positive $V_g$, whereas it increases strongly with temperature under negative gate bias, particularly at larger negative $V_g$. This behavior indicates the thermal activation of hole carriers under negative gating, thereby reaffirming the *p*-type semiconducting nature of 4H-$SnS_2$.

Further, the temperature-dependent transfer characteristics exhibit a pronounced hysteresis in the $I_d$-$V_g$ curves, as highlighted by the red arrows in **Figure** 4(a). The origin of this irreversibility is directly linked to the ferroelectric nature of 4H-$SnS_2$. Under a negative gate voltage, ferroelectric polarization generates bound charges that establish an internal electrostatic field, leading to hole accumulation in the channel. The increased hole density promotes temporary carrier trapping at intrinsic defect states[51], which partially screens the polarization-bound charges. During the reverse gate sweep, the remanent ferroelectric polarization persists, whereas the trapped carriers are released more gradually. Consequently, the drain current remains higher during the backward sweep than during the forward sweep at the same $V_g$, giving rise to the characteristic clockwise hysteresis. The observed hysteresis therefore originates from the coupled evolution of remanent ferroelectric polarization and polarization-assisted charge trapping[49]. Similar hysteretic $I_d$-$V_g$ has been reported in other ferroelectric semiconductors, including $Bi_2O_2Se$[21], $\alpha$-$In_2Se_3$[50], and CIPS[52]. To further investigate the origin of the transfer hysteresis, $I_d$-$V_g$ characteristics were measured at gate-voltage sweep rates ranging from 0.002 to 0.2 V/s (**Figure** S17). The extracted hysteresis-loop area and memory window (MW) are summarized in **Figures** S17(b) and (c), respectively. The hysteresis-loop area exhibits only a modest variation over the sweep-rate range of 0.002-0.08 V/s before decreasing slightly at the highest sweep rate of 0.2 V/s. The MW is defined as the gate-voltage separation between the forward and reverse $I_d$-$V_g$ sweeps at a fixed drain-current level (50% of the maximum drain current), MW = $|V_g^{forward} - V_g^{backward}|$[53]. It remains nearly constant (~ 5-7 V) across the low and intermediate sweep rates and decreases only at 0.2 V/s. Such weak sweep-rate dependence indicates that the dominant switching process occurs on a timescale much shorter than the applied gate-voltage sweep, consistent with intrinsic ferroelectric polarization switching. These results therefore indicate that hysteresis is predominantly governed by ferroelectric polarization, which is also supported by our frequency-dependent PFM measurements (**Figure** S7). These characteristics identify 4H-$SnS_2$ as a promising candidate for nonvolatile memory applications.

Notably, the hysteresis loop exhibits a pronounced non-monotonic temperature dependence, as summarized by the different colored regions in **Figure** 4(c). From $T$ = 80 to 160

K, the hysteresis loop area remains nearly constant (**Figure** 4c; left vertical axis), indicating that the polarization switching is only weakly influenced by thermal activation. Above 160 K, the hysteresis area increases sharply and reaches a maximum around 240 K. Upon further increasing the temperature beyond 240 K, the hysteresis loop area decreases rapidly, coinciding with a sharp rise in the drain current at this temperature (**Figure** 4b), as discussed above. Further, we extracted the MW from the transfer characteristics at each temperature (**Figure** 4c; right vertical axis). It exhibits a pronounced nonmonotonic temperature dependence, varying from ~2.62 to ~6.80 V over the measured temperature range of 80–300 K. The MW increases gradually from low temperatures, reaches a maximum near ~140 K, decreases slowly up to ~220 K, and then drops rapidly between ~220 and 280 K before becoming nearly temperature-independent toward 300 K. Remarkably, this evolution closely follows the temperature dependence of the hysteresis-loop area, indicating that both quantities originate from the same polarization-controlled transport mechanism. The suppression of hysteresis loop area and MW at elevated temperatures is consistent with enhanced screening of the ferroelectric polarization by thermally activated mobile carriers, which weakens the remanent polarization and consequently reduces the transport hysteresis. This highlights the interplay between polarization dynamics and charge transport in the channel.

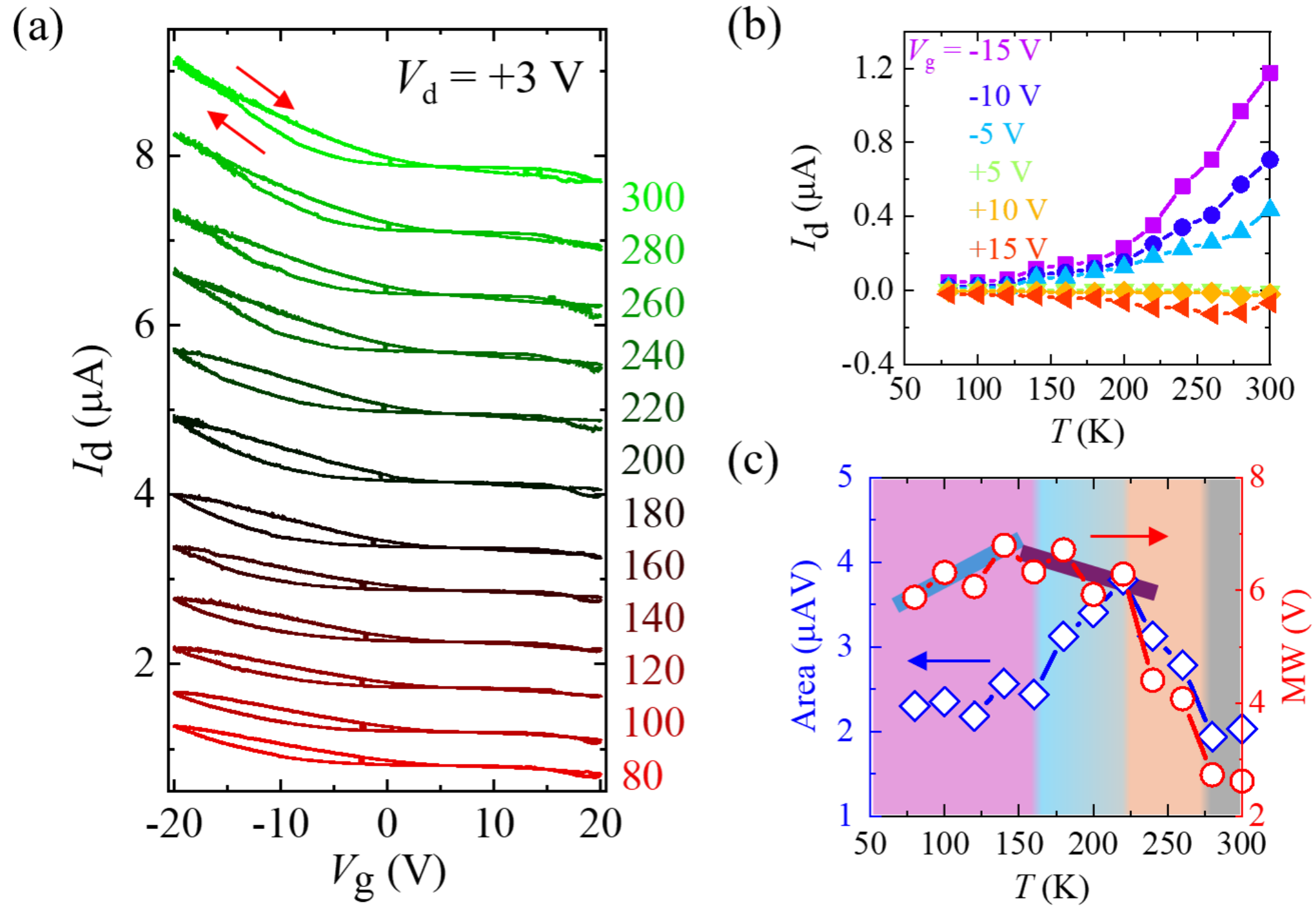

***Figure 4. Temperature and gate-voltage-dependent transport characteristics of the 4H-$SnS_2$ ferroelectric-FET.*** *(a) Transfer characteristics ($I_d$-$V_g$) measured at a fixed drain voltage of $V_d$ = +3 V over the temperature range 80–300 K, illustrating the evolution of channel conduction with temperature; arrows indicate the sweep direction, highlighting the hysteretic behavior. (b) $I_d$ as a function of temperature measured at fixed gate voltages ($V_g$ = -15 V to +15 V), revealing weak temperature dependence at low temperatures and strong thermally activated transport under negative gate bias at elevated temperatures. (c) Temperature dependence of the hysteresis loop area (left vertical axis) and memory window (MW) (right vertical axis) extracted from the transfer characteristics, showing a nonmonotonic evolution with distinct temperature regimes (shaded regions), which reflects the interplay between ferroelectric polarization dynamics and charge transport in 4H-$SnS_2$.*

Since the polarization in 2D ferroelectrics is intimately linked to anharmonic phonon modes, we have investigated the temperature evolution of Raman modes (**Figure** S18) to address the hysteretic behavior of the $I_d$-$V_g$ characteristics. The peak intensities, positions, and linewidths of the $E_g$ and ($A_{1g}$ + $E_g$) were extracted by Lorentzian fitting of the spectra at each temperature. As discussed earlier, the $E_g$ mode features a doublet peak (peaks A and B) at room temperature, which remains visible down to 120 K (**Figure** S18a). This doublet was further deconvolved to extract the peak parameters (**Figure** S18c). As shown in **Figures** S19(a) and (b), the intensity for both modes exhibits a non-monotonic temperature evolution. First, the intensity ratio of peaks A and B in the ($A_{1g}$ + $E_g$) mode remains nearly constant between 120 and 140 K, followed by a pronounced increase above this range and a weak temperature dependence up to ~220 K. Upon further heating, the intensity ratio decreases sharply, reaching a local minimum near ~260 K. Above this temperature, the ratio exhibits a slight recovery and then saturates at higher temperatures. Moreover, the ($A_{1g}$ + $E_g$) mode intensity also exhibits a similar non-monotonic thermal evolution (**Figure** S19b). Such variations in Raman intensity have been attributed to crystal symmetry modifications in vdW metal dichalcogenides[54]. Notably, these phonon anomalies closely track the $I_d$-$V_g$ hysteresis, suggesting a prominent coupling between lattice dynamics and ferroelectric polarization.

The strong lattice-polarization coupling is further manifested in the shifts and FWHM variations of the Raman modes. Specifically, the temperature evolution of the $E_g$ mode exhibits a closer correspondence with the hysteresis behavior. The Raman shift (**Figure** S19c) and FWHM (**Figure** S19e) of both peaks A and B corresponding to this mode display clear anomalies around the characteristic temperatures that separate the different regimes (highlighted by distinct colors in **Figure** 4c) in the hysteresis loop area. On the other hand, the temperature dependence of the peak position for $A_{1g}$+$E_g$ mode reveals a linear redshift from 120 to 300 K (**Figure** S19d), which follows:

$$\omega(T) = \omega_0 + \chi T, \quad (1)$$

where $\omega_0$ denotes the Raman mode frequency at 0 K, and $\chi$ is the first-order temperature coefficient. A linear fit to Eq. 1 yields $\chi \approx (-0.014 \pm 0.001)$ cm$^{-1}\cdot$K$^{-1}$ for the $A_{1g}$+$E_g$ mode, comparable to other 2D materials[55,56], indicating similar interlayer coupling strength. In addition, the FWMH of this mode increases linearly upon heating (**Figure** S19f). While such temperature-dependent Raman shifts and linewidth broadening are consistent with phonon softening driven by lattice thermal expansion and anharmonic phonon-phonon interactions[57], these effects alone cannot account for the non-monotonic temperature dependence of the $I_d$-$V_g$ hysteresis loops presented in **Figure** 4(c). As discussed above, the $E_g$ mode arises from in-plane vibrations of sulfur atoms and is primarily controlled by the Sn-S-Sn bond-bending interactions[32]. The observed anomalies thus suggest temperature-dependent modifications of the local bonding geometry and lattice symmetry within the basal plane, which can influence inversion symmetry breaking and, consequently, ferroelectric polarization. The correlated evolution of the memory window, hysteresis-loop area, and Raman response therefore provides compelling evidence that the observed transport hysteresis is governed predominantly by intrinsic ferroelectric polarization and lattice–polarization coupling, rather than by extrinsic mechanisms such as charge trapping or surface adsorbates.

### 2.3. Light polarization sensitive photoresponse in 4H-$SnS_2$

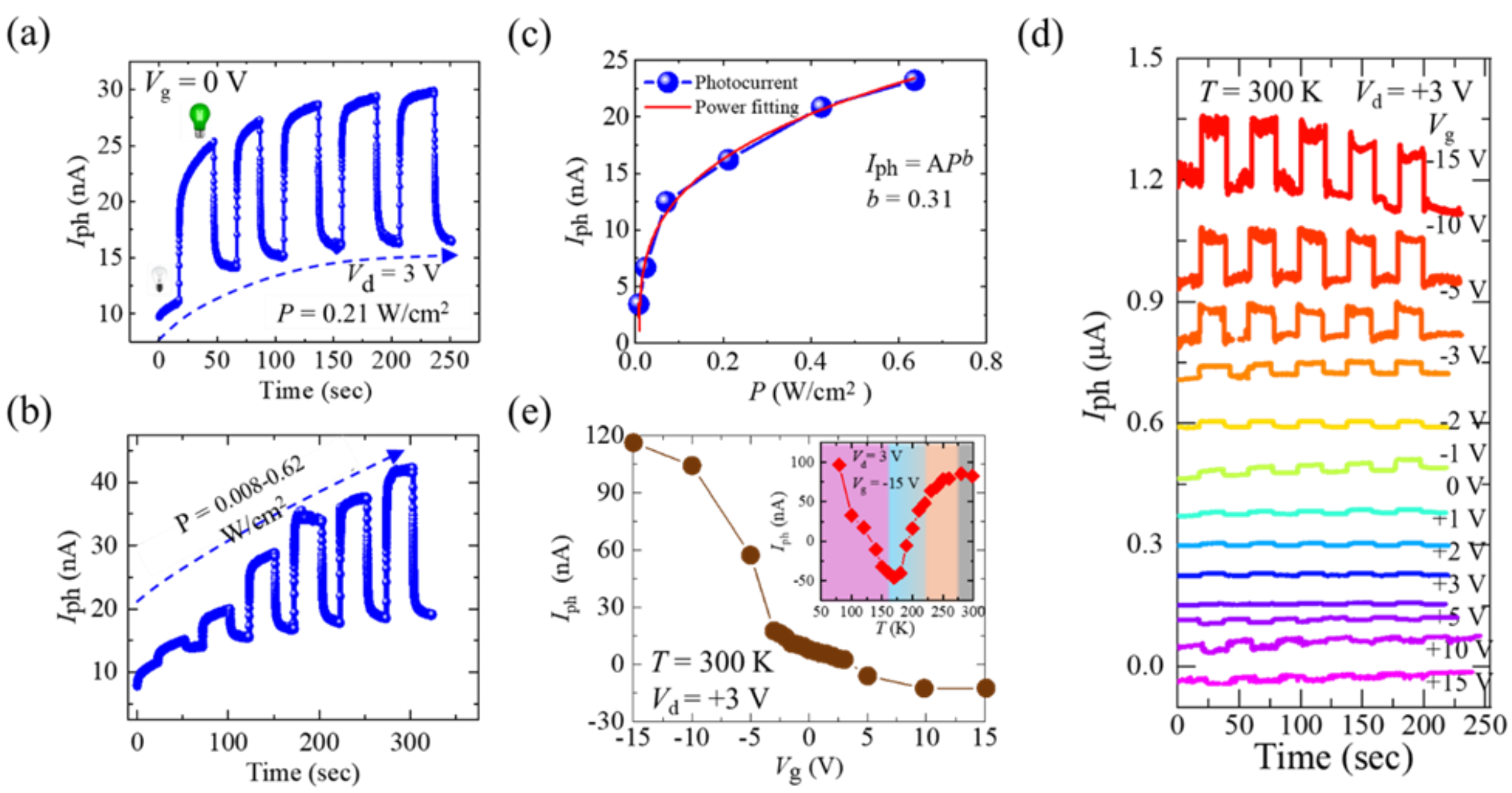

***Figure 5. Photoresponse characteristics of the 4H-$SnS_2$ FET.*** *(a) Time-dependent photocurrent $I_{ph}(t)$ measured at a fixed drain voltage of $V_d$ = +3 V under periodic on–off illumination with a 532 nm laser (0.21 W/cm$^2$), showing reproducible photocurrent switching and step-like current accumulation associated with ferroelectric polarization. The light-bulb icons denote the light ON (green) and OFF (white) states. (b) $I_{ph}(t)$ recorded under systematically increasing laser power from 0.008 to 0.62 W/cm$^2$, illustrating a monotonic enhancement of the photocurrent; the dashed arrow highlights the step-like increase linked to polarization effects. (c) Extracted photocurrent $I_{ph}$ as a function of incident power density, following a power-law dependence $I_{ph} = AP^b$ with b = 0.31 and A = 27.02 nA/ (W/cm$^2$)$^b$, indicative of trap-assisted recombination and polarization screening. (d) Room-temperature $I_{ph}(t)$ measured at a fixed drain voltage of $V_d$ = +3 V under periodic on–off illumination at different $V_g$ ranging from -15 V to +15 V. (e) Extracted $I_{ph}$ as a function of $V_g$. The inset depicts the temperature dependence of $I_{ph}$ measured from 80 to 300 K at $V_d$ = +3 V and $V_g$ = -15 V. The different colored regions show a nonmonotonic evolution with temperature.*

The built-in polarization also influences the photoresponse by controlling the separation of photogenerated carriers. **Figure** 5(a) depicts the time-dependent photocurrent $I_{ph}$(t) at a fixed $V_d$ (+3 V) under 532 nm laser illumination with a power of 0.21 W/cm$^2$. Upon illumination, the source–drain current increases rapidly, indicating a strong photocurrent response. Five successive on–off cycles exhibit highly reproducible current profiles at room temperature. This photoresponse is assisted by the internal electric field associated with ferroelectric polarization, promoting efficient separation of photogenerated electron-hole pairs in the channel. An interesting feature of the data shown in **Figure** 5(a) is the gradual increase in dark current on successive illumination cycles as shown by the dotted lines. This is suggestive of a change in the polarization state of the material which affects the barrier height at the metal-$SnS_2$ interface. **Figure** 5(b) presents the variation of $I_{ph}$(t) measured under increasing laser power from 0.008 to 0.62 W/cm$^2$. The extracted photocurrent ($I_{ph} = I^{light}{}_{ph} - I^{dark}{}_{ph}$) increases monotonically with power, reflecting enhanced photocarrier generation at higher photon flux. Here also, we see a gradual rise in dark current after successive illuminations of increasing power. The power density dependence of $I_{ph}$, shown in **Figure** 5(c), follows power-law dependence,

$$I_{ph} = \mathrm{A}P^{b}, \quad (2)$$

where the prefactor A and exponent b are 27.02 nA/ (W/cm$^2$)$^b$ and 0.31, respectively. This sublinear dependence on power ($b < 1$) suggests a complex behavior of carrier generation, trap states, recombination process, and polarization screening effects. To further evaluate the optoelectronic performance, the responsivity ($R$), specific detectivity ($D$), and external quantum efficiency (EQE) were calculated and are shown in Figure S20. The responsivity was obtained from $R = I_{ph}/ (P\mathrm{A})$, where $I_{ph}$ is the photocurrent, $P$ is the incident optical power density, and A is the illuminated device area. The specific detectivity was then obtained assuming shot-noise-limited performance, $D = \frac{R\sqrt{S}}{\sqrt{2qI_d}}$, where $I_d$ is the dark current and $q$ is the elementary charge. The maximum responsivity and specific detectivity obtained for the present 4H-$SnS_2$ device are approximately ~0.6 nA/W and ~$3\times10^8$ Jones, respectively. While the responsivity is lower than that reported for 2D ferroelectrics such as $CuCrP_2S_6$[58] and $CuInP_2S_6$[59], the detectivity remains of the same order of magnitude. Finally, the external quantum efficiency was calculated from EQE = $R\,\frac{\mathrm{hc}}{q\lambda}$ , where $h$ is Planck's constant, $c$ is the speed of light, and $\lambda$ is the excitation wavelength[60]. The relatively high values obtained at low illumination intensity indicate efficient photocarrier generation and collection in the device. As the illumination power increases, $R$, $D$, and EQE decrease monotonically. This behavior is consistent with a sublinear photocurrent response $I_{ph} = \mathrm{A}P^{b}$ ($b < 1$). Further, the correlation between photoresponse and ferroelectric hysteresis is also manifested in the temperature-dependent photocurrent measurements under constant and varying laser powers, similar to that of $I_{ph}$(t) data presented in **Figures** 5(a) and (b). As shown in **Figure** S21, the $I_{ph}$(t) at a fixed $V_d$ (+3 V) and power of 0.21 W/cm$^2$ displays a non-monotonic evolution with temperature, with a minimum value at 190 K. Additionally, the $I_{ph}$(t) at different powers ranging from 0.008-0.62 W/cm$^2$ also show similar behavior, again featuring minima at 190 K (**Figure** S22). Such behavior directly links the photoresponse with ferroelectric polarization dynamics in 4H-$SnS_2$, suggesting that the illumination acts not only as a carrier source but also as an effective stimulus for the time dependence of polarization which effectively leads to a charge current in the system.

**Figure** 5(d) shows the gate voltage ($V_g$) dependence of $I_{ph}$(t) when $V_g$ is changed between +15 to -15 Volts while keeping the $V_d$ fixed at +3V. As summarized in **Figure** 5(e), $I_{ph}$ exhibits a strongly gate-tunable response. At negative $V_g$, the ferroelectric polarization field associated with the built-in spontaneous polarization enhances carrier separation and transport of photogenerated carriers, leading to a steady rise in $I_{ph}$ up to $V_g = -3$ V, followed by a sharp increment to a large positive $I_{ph}$ value (~120 nA). In contrast, applying a positive $V_g$ modifies the polarization configuration and electrostatic potential, effectively reversing the direction of the net polarization-assisted internal field relative to the carrier flow. This suppresses carrier separation and causes $I_{ph}$ to decrease gradually to a small negative value of nearly −15 nA. The behavior of $I_{ph}$ under negative $V_g$ is consistent with in the increased carrier density in a *p*-type channel. However, the abrupt increase in $I_{ph}$ beyond $V_g \leq -3$V suggests an additional ferroelectric-assisted photoconductive response, where polarization further controls photocarrier separation and transport[21]. Similar polarization-enhanced photoresponse has been reported in 2D ferroelectrics $NbOCl_2$[61]. This coupling is further reflected in the temperature-dependent photoresponse measurements. The inset of **Figure** 5(e) shows the thermal evolution of $I_{ph}$ measured at $V_d = +3$ V and $V_g = -15$ V. At 80 K, the photocurrent is positive and large (~100 nA). As the temperature increases, $I_{ph}$ drops steadily and becomes negative before reaching a minimum near ~160 K. Upon further warming, the photocurrent rises and recovers to a large positive value at ~220 K, and beyond this temperature a weak temperature dependence of the $I_{ph}$ is seen. Interestingly, such unusual temperature dependence of $I_{ph}$ correlates with the evolution of ferroelectric hysteresis loop in $I_d$ vs $V_g$ curves acquired under similar conditions (**Figure** 4c), as well as with anomalies

observed in the Raman response (**Figure** S18). This correlated evolution of Raman features, $I_d$-$V_g$ hysteresis, and photoresponse thus suggests that the non-monotonic thermal evolution arises from competing temperature-dependent processes in photocarrier generation, recombination, polarization, and transport, rather than from purely thermal anharmonic effects.

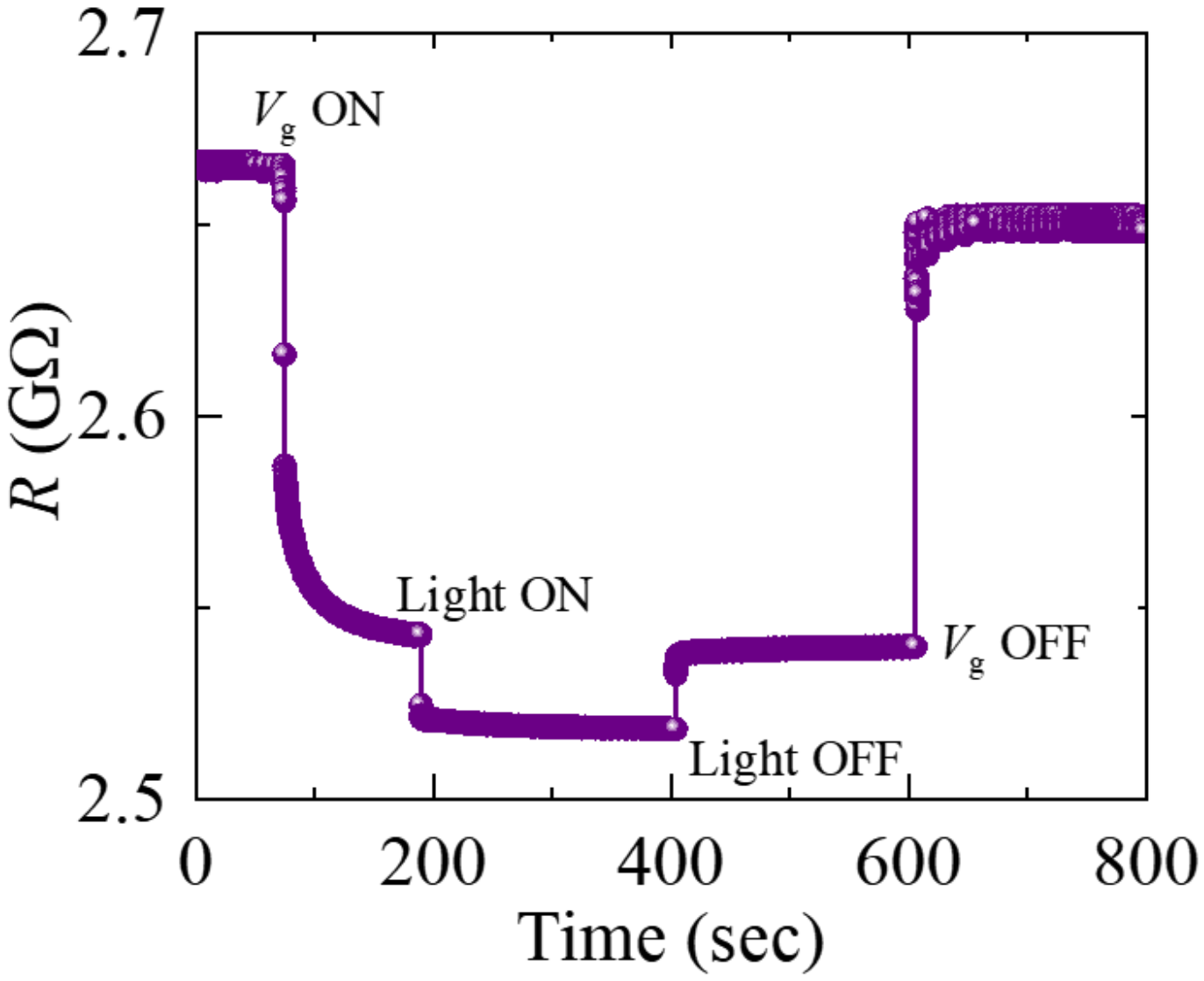

***Figure 6.*** *The effects of sequential application of gate voltage ($V_g$ = –15 V) and exposure to light (with the same flux as in the light-only experiment) on the resistance of the channel of 4H-$SnS_2$.*

In **Figure** 6, we illustrate the complementary roles of gate voltage and photoillumination in modulating the channel conductivity. Upon applying a negative gate voltage of $V_g = -15$ V, the resistance of the device decreases abruptly from ~2.66 to ~2.55 GΩ, followed by a slower relaxation until exposure to light. Subsequent illumination with a photon flux equal to the one used in the light-only experiment in **Figure** 5 further enhances conductivity, leading to an additional reduction in resistance to nearly 2.52 GΩ. When the light is switched off, the resistance recovers only partially and remains nearly constant in time. Almost full recovery to the initial high-resistance state occurs only after removing the gate voltage, and this process is instantaneous. This behavior persists down to temperatures as low as 100 K (**Figure** S23). This feature of photoresponse demonstrates that the electric field and light cooperatively control the conductive state, enabling tunable optoelectronic functionality.

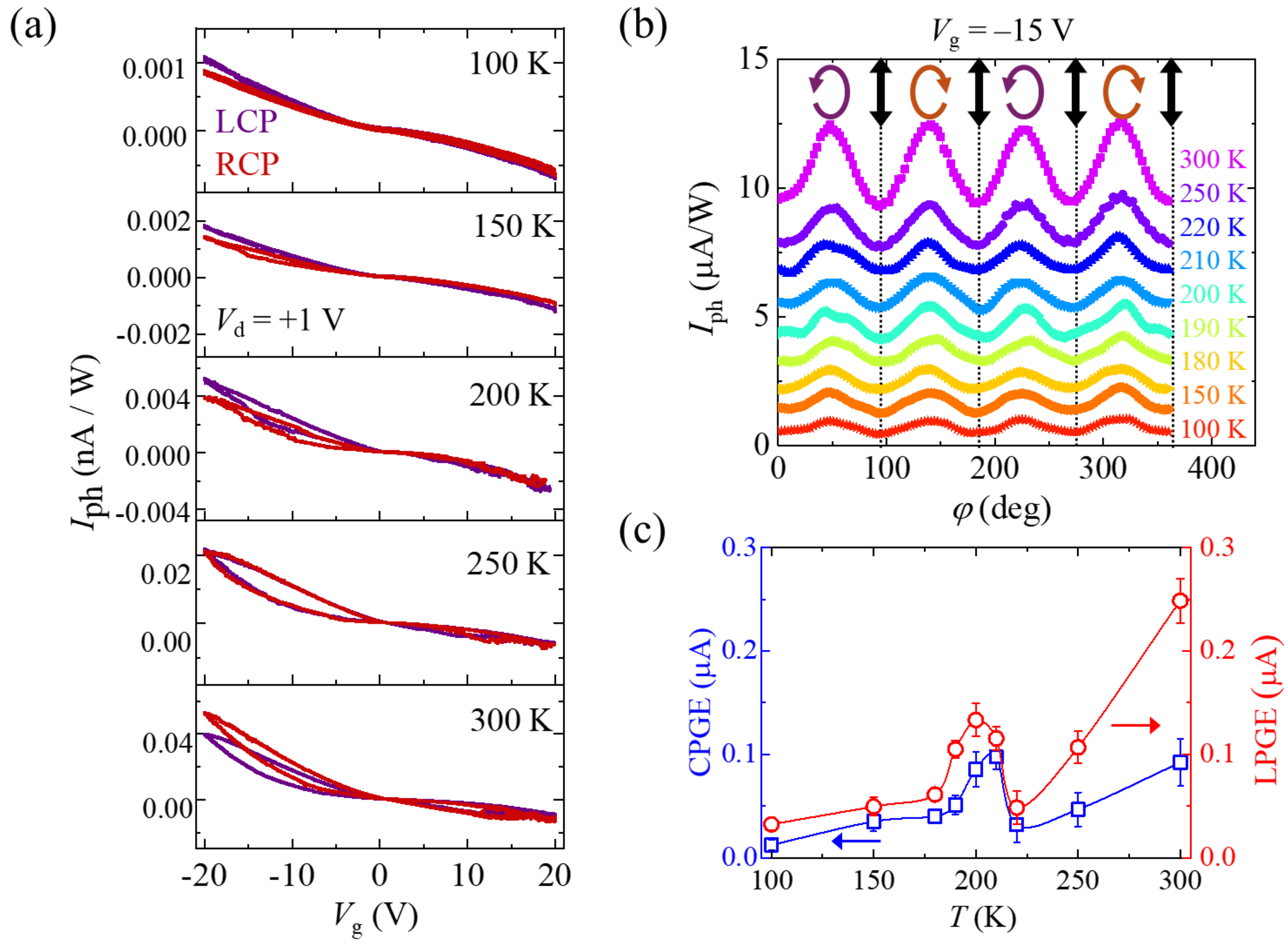

***Figure 7. Circular photogalvanic effect of the 4H-SnS₂.*** *(a) Transfer characteristics ($I_{ph}$-$V_g$) measured at a fixed drain voltage of $V_d$ = +1 V over the temperature range 100–300 K under LCP and RCP light. (b) The polarization-dependent photocurrent as a function of QWP angle φ at different temperature measured at fixed gate voltage ($V_g$ = -15 V). (c) Temperature dependence of CPGE and LPGE amplitudes extracted from fitting $I_{ph}(\varphi)$ in* ***Figure*** *6(b) using the expression: $I(\varphi) = D + C\sin(2\varphi) + L_1 \sin(4\varphi) + L_2 \cos(4\varphi)$.*

The inherent symmetry breaking associated with spontaneous polarization provides a potential platform for the observation of circular photogalvanic effects (CPGE)[62,63]. In such a scenario, circularly polarized light could generate helicity-dependent photocurrents through asymmetric carrier excitation. To elucidate this, we performed the gate-dependent photoresponse under left- (LCP) and right-circularly polarized (RCP) light. **Figure** 7(a) compares the $I_{ph}$ as a function of $V_g$ measured from $T$ = 100 K to 300 K at a fixed $V_d$ = +1 V under the LCP and RCP conditions. The distinct separation between the two responses demonstrates a clear helicity-dependent photocurrent over the entire gate sweep. The difference between LCP and RCP is most pronounced under negative $V_g$ and becomes progressively weaker as $V_g$ moves to positive values. The enhanced contrast at negative $V_g$ suggests that the carrier population or band alignment couples more effectively to circularly polarized excitation, whereas positive gate voltage suppresses this asymmetry. The helicity-dependent response is also affected by temperature. At low temperatures ($T \leq 200$ K) the separation between the LCP and RCP curves is apparent. However, on further heating to 250 K, the two curves nearly overlap for the entire gate sweep. Notably, above 250 K, the relative sign of the asymmetry between LCP and RCP reverses, indicating a crossover in the dominant helicity-dependent carrier transport at 300 K. Interestingly, this characteristic temperature (~250 K) where a crossover takes place coincides with the peak in ferroelectric hysteresis observed in **Figure** 4(c). Such behavior emphasizes the role of polarization in governing the helicity-sensitive photoresponse.

The coupling between ferroelectric polarization and helicity-dependent photocurrent is further revealed on rotating the quarter-wave plate (QWP) from $f = 0$ to 360 degrees with respect to the polarization axis of the incident beam (**Figure** 7b). This rotation modulates the light polarization periodically (180°), evolving from linear polarization (LP) at $f = 0°$, to RCP at $f = 45°$ back to LP at $f = 90°$, to LCP at $f = 135°$, and again to LP at $f = 180°$. **Figure** 7(b) shows the behavior of $I_{ph}$ $(f)$ at several temperatures with $V_g = -15$ V, where a distinct periodicity of p/2 is seen. The polarization-dependent photocurrent as a function of QWP angle $f$ can be described as[64]:

$$I(\varphi) = D + C \sin(2f) + L_1 \sin(4f) + L_2 \cos(4f), \quad (3)$$

where the term with coefficient $C$ represents the CP-sensitive $I(f)$ with a 180° periodicity. The $L_1$ and $L_2$ terms correspond to the LP-dependent contributions with a 90° periodicity that denote the magnitude of the linear photogalvanic effect (LPGE). The $D$ term in Eq. 3 represents the polarization-independent photocurrent. The $I(f)$ plot of **Figure** 7(b) is fitted to Eq. 3, which yields the amplitudes of the CPGE and LPGE contributions to $I(f)$ plotted as a function of temperature in **Figure** 7(c). Both CPGE and LPGE exhibit a similar non-monotonic temperature dependence, which increase to a maximum around 200–220 K, followed by a suppression and subsequent enhancement at higher temperatures. Notably, the maxima in CPGE and LPGE closely coincide with the peak in ferroelectric hysteresis near this temperature, indicating a strong correlation between photoinduced carrier and ferroelectric polarization dynamics. The observed polarization dependence originates from inversion-symmetry breaking inherent to the polar 4H-$SnS_2$ phase.

Such broken symmetry enables asymmetric optical excitation and nonequilibrium carrier generation under circularly and linearly polarized light, highlighting that the ferroelectric polarization directly controls the photocarrier dynamics.

## 3. Conclusion

In summary, we establish a direct correlation between structural polytype and ferroelectric functionality in $SnS_2$ by systematically comparing the 2H and 4H phases. We further show that the polar 4H phase enables electrical- and light-tunable transistor behavior in a three-terminal FET device. The coexistence of 2H and 4H-$SnS_2$ polymorphs is resolved through complementary structural and vibrational characterization, where X-ray diffraction confirms the layered crystal structure and Raman spectroscopy reveals distinct phonon signatures unique to each phase. Room-temperature ferroelectricity is observed in the 4H phase through PFM phase hysteresis and butterfly-shaped amplitude loops, whereas the 2H polymorph exhibits no detectable ferroelectric response. Electrical measurements reveal polarization-dependent output characteristics, gate-tunable hysteresis, and temperature-driven carrier activation in 4H-$SnS_2$. The photoresponse further exhibits polarization-assisted carrier separation, power-law scaling, and a nonmonotonic temperature dependence that correlates closely with ferroelectric hysteresis dynamics. These findings establish the synergistic interplay between spontaneous polarization, mobile carriers, and photoresponse, establishing 4H-$SnS_2$ as a promising platform for polarization-controlled electronics and optoelectronics for nonvolatile memory and photoferroelectric devices.

## 4. Experimental Section

*Crystal growth and characterization:* The single crystals of 2H- and 4H-$SnS_2$ were synthesized from the stoichiometric mixture (~2 g) of elemental ingredients using the chemical vapor transport (CVT) method with iodine (~40 mg) as the transport agent. The ~10 $mm^2$ area light-green (G) and wine-red crystals (R), corresponding to the 2H and 4H phases, respectively, were simultaneously obtained within the same growth ampoule after 1 week of CVT growth with a temperature gradient from 850 to 600°C.

The elemental compositions of both sets of $SnS_2$ crystals were examined by energy-dispersive X-ray spectroscopy (EDS). These crystals were also characterized by X-ray diffraction (XRD),

Raman spectroscopy, and photoluminescence (PL) measurements. The XRD was performed using a Rigaku miniflex diffractometer with Cu-$K\alpha1$ radiation. A typical diffraction pattern of the crystals placed such that the scattering vector is parallel to the c-axis has only the (00$L$) reflections (**Figure** 1a). The Raman and PL measurements were conducted with an XploRA PLUS confocal Raman microscope (Horiba) with 532 and 405 nm laser excitation, respectively.

The surface topography, thickness, ferroelectricity, polarization switching, and local surface potential of exfoliated $SnS_2$ nanoflakes placed on a Pt-coated Si substrate were studied using the NX10, Park Systems scanning probe microscope at room temperature. The ferroelectric and piezoelectric properties were characterized by PFM using a Pt/Cr-coated conductive tip (spring constant: 3 N $m^{-1}$; tip sensitivity: 33.33 V $\mu m^{-1}$). Domain-switching behavior was investigated by box-in-box poling experiments acquired in off-resonance PFM (OR-PFM) mode at 17 kHz with an AC drive voltage of 1 V. Local piezoresponse hysteresis measurements, including phase and amplitude loops, were subsequently performed using the same tip in contact-resonance PFM (CR-PFM) mode. Surface potential mapping was carried out by KPFM using an Au/Cr-coated conductive AFM probe.

*Electrical measurement:* To probe the electronic properties of $SnS_2$, we fabricated three-terminal phototransistors directly on bulk 4H-$SnS_2$ single crystals (~100 µm thick). Silver (Ag) electrodes deposited on the top surface of the crystal served as the source and drain contacts with a channel spacing of ~25 µm, while a patterned Ag pad beneath the crystal functioned as the gate electrode, as shown in the cross-sectional schematic of the device (**Figure 3b**). Although Ag electrodes are known to undergo electrochemical migration under sufficiently high electric fields at room temperature, the nearly linear output characteristics observed over the measured drain-bias range (**Figure** 3) show no abrupt "SET/RESET" transitions or large resistance changes that are characteristic of Ag filament formation in electrochemical metallization memories[65]. In this architecture, the 4H-$SnS_2$ single crystal simultaneously serves as the semiconducting channel and the ferroelectric gate dielectric. Consequently, the applied gate voltage establishes an electric field through the thickness of the crystal, enabling direct electrostatic modulation of the channel conductivity. Electrical measurements were performed using a two-channel sourcemeter (Keithley 2602B), enabling precise control and measurement of voltage and current signals[66,67]. The electrical transport anisotropy of the 4H-$SnS_2$ crystal was characterized by measuring the in-plane and out-of-plane resistances. The in-plane resistance was measured between the source and drain electrodes, whereas the out-of-plane resistance was measured between the source (or drain) and the bottom gate electrode. A constant DC voltage was applied, and the resulting current was recorded to determine the resistance. The corresponding in-plane and out-of-plane resistivities were calculated using the measured device geometry (**Figure** 2b). Time-dependent relaxation measurements were performed by applying a programming pulse for a fixed duration to program the device, followed by recording the drain current as a function of time at a constant drain voltage after removing (or resetting) the programming pulse. The current was monitored for several minutes to evaluate the temporal stability of the programmed state. For polarization-dependent transport measurements, an erase-program-read protocol was employed to ensure a well-defined ferroelectric polarization state prior to each measurement. Before each programming operation, the gate was initialized using an erase pulse of opposite polarity to the subsequent programming pulse. The erase-pulse amplitude was chosen to be 2 V larger than the corresponding programming voltage (e.g., +10 V before −8 V programming, +8 V before −6 V programming, and −10 V before +8 V programming), with a pulse duration of 100 ms. Subsequently, a programming voltage ($V_p$) ranging from −10 to +10 V (pulse duration: 100 ms) was applied to the gate electrode. After the programming pulse, the gate bias was removed ($V_g$ = V), and the output characteristics ($I_d$-$V_d$) were immediately measured. This erase-program-read cycle was repeated independently for each programming voltage, thereby eliminating hysteresis-induced history effects and ensuring that each output characteristic corresponds to a well-defined ferroelectric polarization state.

*Photoconductivity measurement:* For the photoexcitation measurements, we used a 532 nm laser beam passing through a linear polarizer, followed by a quarter-wave plate and then reflecting off a dielectric mirror and impinging on the sample surface at an angle $\theta$, as sketched in **Figure 3**a. For normal photocurrent measurements, the laser beam is normal ($\theta$ = 90°) to the sample surface, whereas it strikes at an angle $\theta$ = 45° in light helicity-dependent experiment. The laser power at the sample location was measured with a precision power meter. The areal density of power varied from 0.008 to 0.62 W/cm$^2$. All electrical transport and photoconductivity measurements were carried out under high-vacuum conditions inside a liquid nitrogen flow cryostat equipped with a quartz optical window (Linkam Scientific Instruments; Serial Number: 11097-0148) to suppress the influence of ambient adsorbates and enable reliable temperature-dependent characterization.

**Supporting Information**

Supporting Information is available from the Wiley Online Library or from the author.

**Acknowledgements**

A.B., R.B., and S.C. contributed equally to this work. It is funded by the Department of Defense through grant # W911NF2120213 to the DOD Center of Excellence for Advanced Electro-Photonics with Two-Dimensional Materials at Morgan State University.

**Data Availability Statement**

The data that support the findings of this study are available from the corresponding author upon reasonable request.